# mm-Wave Low Cost MIMO Antennas with Beam Switching Capabilities Fabricated Using 3D Printing for 5G Communication Systems

Shaker Alkaraki and Yue Gao

*Abstract—* **This paper presents designs and prototypes of low cost multiple input multiple output (MIMO) antennas for 5G and millimetre-wave (mm-wave) applications. The proposed MIMOs are fabricated using 3D printing and are able deliver beams in multiple directions that provide continuous and real time coverage in the elevation of up to ∓ 30° without using phase shifters. This equips the proposed MIMO with a superior advantage of being an attractive low cost technology for 5G and mm-wave applications. The proposed MIMO antennas operate at the 28 GHz 5G band, with wide bandwidth performance exceeds 4 GHz and with beam switching ability of up to ∓ 30° in the elevation plane. The direction of the main beam of the single element antenna in the MIMO is steered over the entire bandwidth through introducing a 3D printed walls with different heights on the side of the 3D printed radiating antenna. Unlike all other available beam steering techniques; the proposed wall is not only able to change the direction of the beam of the antenna, but also it is able to increase the overall directivity and gain of the proposed antenna and MIMO at the same time over the entire bandwidth.**

## I. INTRODUCTION

The upcoming fifth generation (5G) is a wireless mobile communication technology that is expected to be completely standardized by 2020 and will satisfy the ever-growing demand on wireless data transmission. For example, 5G is expected to improve the overall system capacity by several hundred times than the existing technologies and to increase the overall system throughput with higher spectral and energy efficiency, while minimizing the system latency [1-2]. Unlike its predecessors, 5G will be introduced over several frequency bands. Lower frequency bands which are sub 6 GHz are heavily populated, and have limited bandwidth available. Hence, to meet 5G requirements, several frequency bands at millimetre-wave (mm-wave) have been proposed and approved in most of the developed countries. This includes (24 GHz to 29.5 GHz), (37 GHz to 42.5 GHz), (47.2 GHz to 48.2 GHz) and (64 to 71 GHz) [2-4].

Three dimensional printing (3D printing) also known as additive manufacturing is an effective technology that have been widely used to deliver efficient designs for various antenna applications at different frequency bands ranges from microwave to THz frequencies. Using 3D printing to provide antenna solutions have several advantages such as realizing complex shapes at low cost especially, if combined with low cost metallization techniques as shown in [5]. Generally, manufacturing an antenna using 3D printing involves two major steps: the first is to 3D print the antenna using a 3D printer and then to metallize the 3D printed structure using a proper metallization technique. This metallization technique can be the well-known and relatively high cost electroless plating used in [6-7] or a low cost metallization technique such as the techniques recently proposed in [5].

Multiple input multiple output (MIMO) system was developed by Bell lab and since then it has become a major component of wireless communication system. In a MIMO system, multiple antenna are deployed on both transmitter and receiver in which will increase the system capacity linearly with increasing the number of receive and transmit antennas [8-10]. A basic MIMO system consists of an array of M transmit antennas that transmits signal to K antennas at the receiver side where each antenna in both sides has its different and independent RF chain. MIMO adds another dimension to the capacity *C* of the communication link in comparison of single input single output (SISO) systems as the capacity of the a MIMO link can for a given signal to noise to ratio (SNR) can be expressed in bits/s as following [8-10]:

$$C = B \min(M,K) \ (\log_2(1 + SNR)) \qquad (1)$$

Where, *B* is the bandwidth of the system expressed in Hz.

Establishing a reliable communication link at mm-wave bands for cellular applications offers great opportunities and poses several challenges. Mm-wave bands provide wider bandwidth, compact antenna size and miniaturized dimensions. However, the main challenges arise from moving to mm-wave bands are: high atmospheric signal losses, shadowing and high cost of system components [11]. The attenuation of the signal at mm-wave mainly depends on the propagation distance, weather conditions and operating frequency. Shadowing is another important source of signal losses [12]. These losses introduces a challenge for antenna engineers to develop efficient, steerable and high gain antennas that help to overcome these losses to establish high quality communication links at mm-wave frequencies. Hence, 5G mm-wave antenna base stations are expected to use a combination of beamforming and MIMO technology. Beamforming is needed to provide beam steering abilities as well as to provide highly directive beam towards mobile phone users. While, MIMO technology can be used to increase the overall capacity of the system by increasing the data rates by transmitting several data streams to a user which is referred as single MIMO (SU-MIMO) and by serving

This work is supported by Physical Sciences Research Council (EPSRC) in the U.K. under Grant EP/R00711X/1. Shaker Alkaraki and Yue Gao are with the School of Electronic Engineering and Computer Science, Queen Mary University of London, London, E1 4NS, U.K. (e-mail: s.m.alkaraki, yue.gao@qmul.ac.uk).

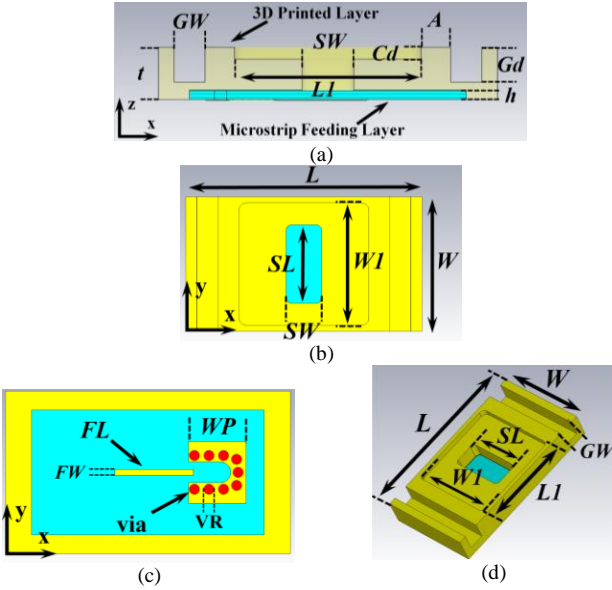

Fig.1. The schematic of the proposed single element antenna. (a) Cross section of front view, (b) top view, (c) bottom view, and (d) perspective view.

multiple users simultaneously which is referred as MU-MIMO [13].

Realizing antennas and RF front-end solutions for 5G mm-wave base stations that are able to deliver both high gain and beam-steering functionalities are highly complex and expensive task due to the high cost of elements used in the system design such as phase shifters. For example, phased arrays are extensively used in different applications to deliver high gain and efficient steerable beam at low and high frequency. However, the high cost of realizing phased arrays, especially at mm-wave remains one of the main challenges for the phased array technology to be deployed in future 5G mm-wave base stations.

In this paper, we propose an innovative and low cost MIMO antenna for 5G mm-wave base station applications. The proposed MIMO antenna is fabricated using 3D printing, which offers the opportunity to deliver innovative and complex antenna designs with an overall reduced cost in comparison to the conventional antennas. The proposed MIMO antenna is compact, low cost, efficient, high gain and it provides beam-switching abilities using a novel technique without using phased array technology.

This paper is organized as follows. Firstly, the structures and prototypes of the proposed antenna, $2 \times 2$ MIMO and $4 \times 3$ MIMO with beam switching abilities are shown. Secondly, the fabrication method and the operation principles of the single element antenna that is used to build the proposed MIMOs are explained in details. Then, the paper explains the beam steering mechanism which has been implemented to steer the beam of the antenna through introducing a 3D printed metallized wall on the side of the antenna. Finally, the measured and simulated results of the proposed MIMO antennas are presented as well as the diversity analysis of the performance of the proposed MIMO antennas.

## II. ANTENNA STRUCTURE AND FABRICATION METHOD

The structures and prototypes of the proposed antenna, MIMO and steer-able MIMO are shown in this section. This section is organized as following: firstly, the structure of the single element antenna that is used to realize the proposed MIMO antennas is presented. Then, the structure of a $2 \times 2$ MIMO which is designed based on the single element antenna is shown. After that, the structure of the single element antenna combined with sidewall which is used to steer the beam of the antenna is shown. Then, the structure and the prototype of the proposed $4 \times 3$ steer-able MIMO are presented.

### A. The Design of Single Element Antenna

The structure of the proposed single element antenna is shown in Fig.1. The antenna is designed based on the design principles presented in [14-18]. The proposed antenna consists of two main different structures as shown in Fig.1, which are the feeding structure and the radiating structure. The feeding structure is a microstrip feeding layer consists of mini-smp ground plane/pad, vias and transmission line fabricated using RO4003C substrate with a dielectric constant of 3.38. The feeding layer is designed to couple the electromagnetic energy to the surface of the radiating structure using transmission line and using PE44968/PE44489 mini-smp connector which are placed on the bottom of the feeding layer on the ground plane as shown in in Fig.1 (c) and (d). However, the radiating structure is the 3D printed element, which consists of a central slot surrounded by a rectangular cavity and two corrugations. The dimensions of the antenna are summarized in Table.1 and are described as following: length of antenna ($L$), width of antenna ($W$), thickness ($t$), width of the slot ($SW$), length of slot ($SL$), feed line width ($FW$), length of feed line ($FL$), top cavity depth ($Cd$), top cavity width ($W1$), top cavity length ($L1$), radius of via ($VR$), width of corrugations ($GW$), depth of corrugations ($Gd$), distance between cavity and corrugation ($A$) and width of mini-smp pad ($WP$).

Table1. The dimension of the proposed single element antenna.

| Parameter | Dimensions | Parameter | Dimensions |
|---|---|---|---|
| W | 11 mm | GW | 1.8 mm |
| L | 19.6 mm | Gd | 2 mm |
| SL | 6.5 mm | VR | 0.7 mm |
| SW | 3 mm | FW | 0.4 mm |
| W1 | 10 mm | FL | 5.4 mm |
| L1 | 10.8 mm | A | 1.7 mm |
| Cd | 0.7 mm | WP | 4.2 mm |
| t | 3 mm | h | 0.508 mm |

### B. Steerable Single Element with Side Wall

The introduction of a metallized 3D printed wall on the side of the single element antenna as shown in Fig.2, will steer the beam of the antenna to the opposite direction in the (y-z) plane and it will improve the overall gain of the antenna. The beam steer-ability is proportional to the wall height ($Wh$) till reaching saturation limit of ≈30°.

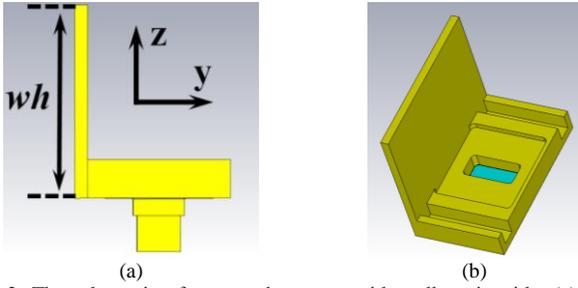

Fig.2. The schematic of proposed antenna with wall on its side. (a) Front view and (b) perspective view.

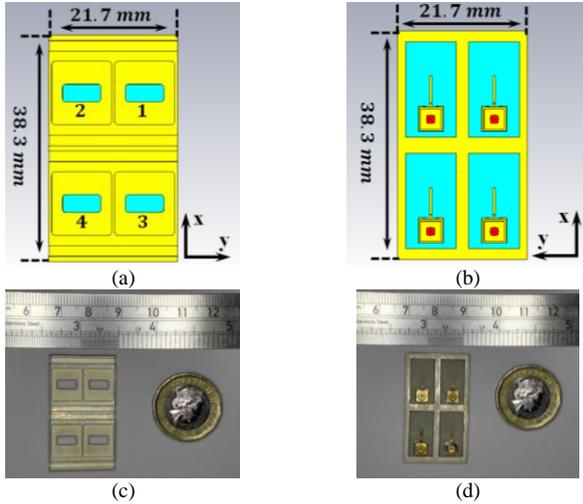

Fig.3. The schematic and prototype of the proposed $2 \times 2$ MIMO antenna. (a) Schematic of top view, (b) schematic of bottom view, (c) top view of the fabricated prototype and (d) bottom view of the fabricated prototype.

## C. $2 \times 2$ MIMO and $4 \times 3$ MIMO with Beam Switching Capabilities

The structure of the proposed $2 \times 2$ MIMO is shown in Fig.3 (a) and (b). The MIMO consists of four elements that have an identical size and dimensions to the single element antenna shown in Fig.1. The overall dimensions of the MIMO is $21.7\ mm \times 38.3\ mm$ and the top and bottom view of the fabricated prototype are shown in Fig.3 (c) and (d). Furthermore, the schematics of the $4 \times 3$ steerable MIMO that is able to provide continuous beam coverage up to $\pm \approx 30°$ in the elevation as well as radiation in the boresight direction are shown in Fig.4. The MIMO consists of 12 elements as shown in Fig.4, where six elements (element 1-6) are designed with different wall height and they are designed to provide steerable beams in the direction of $\pm \approx 10°$, $\pm \approx 20°$ and $\pm \approx 30°$. The other six elements (element 7-12) are identical to the single element antenna with no sidewall as they radiate in the boresight direction. The total dimensions of the $4 \times 3$ MIMO are $52.9\ mm \times 57.0\ mm$ and the prototype of the design is shown in Fig.4. The 6 elements which radiate in the boresight are $10\ mm$ ($\approx \lambda$ at 28 GHz) away from the elements with the wall to eliminate the effect of the walls on the radiation patterns of the antennas that radiate in the boresight direction.

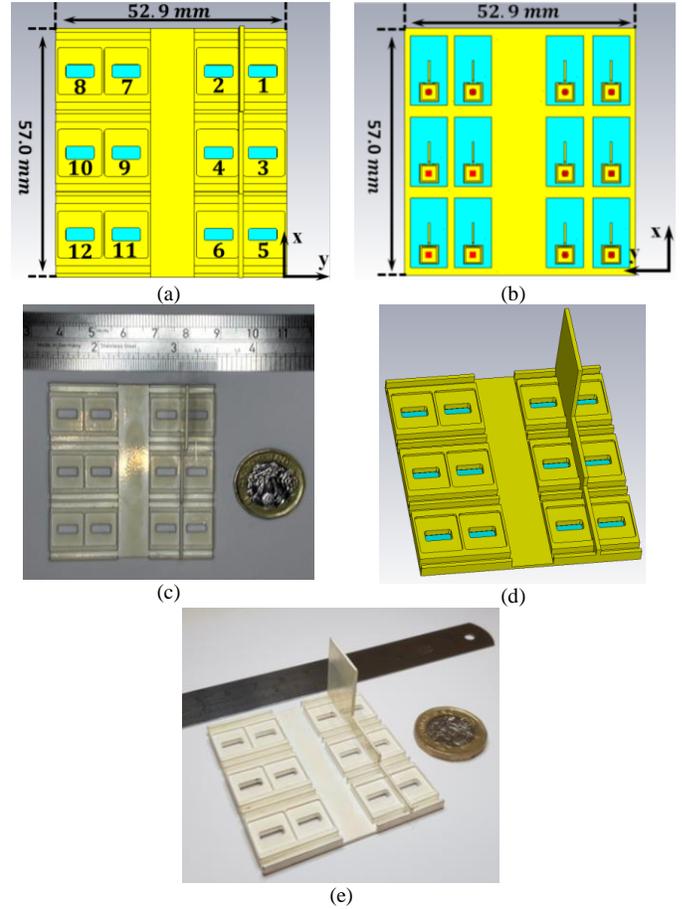

Fig.4. The schematics and the prototype of the proposed $4 \times 3$ MIMO with side walls. (a) Schematic of top view, (b) schematic of bottom view, (c) top view of the fabricated prototype, (d) schematic of perspective view and (e) perspective view of the fabricated prototype.

## D. Fabrication Method

The proposed antennas and MIMO consists of two different structures that are fabricated separately and then assembled to produce the final design. The first structure is the feeding microstrip layer which is fabricated using conventional PCB fabrication method. The second structure is the radiating structure which is fabricated using 3D printing as following: firstly, the structure is 3D printed with a clear finish using Objet30 3D printer that prints with a layer thickness of $16\ \mu m$ and a resolution of $100\ \mu m$ using verlo white material. Then, the 3D printed structure is rinsed using pressurized water to remove the support material which has been used during the 3D printing process. After that, the antenna is metallized using Jet Metal (JMT) Process that described in details in [5]. JMT process involves spray coating the 3D printed plastic structure with a thin and very smooth layer of silver that has a conductivity of $4.4 \times 10^7\ s/m$ and thickness of $2.5\ \mu m$. This is significantly thicker than the skin depth at 28 GHz which is $\sim 0.4\ \mu m$ [5]. Finally, using 3D printing and JMT process for metallizing the radiating layer reduces the overall weight and cost of the proposed MIMO. For example, the overall weight of the 3D printed $4 \times 3$ steerable MIMO is less than 20 grams with an overall cost of few dollars for the material used for 3D printing and metallizing the prototype [5].

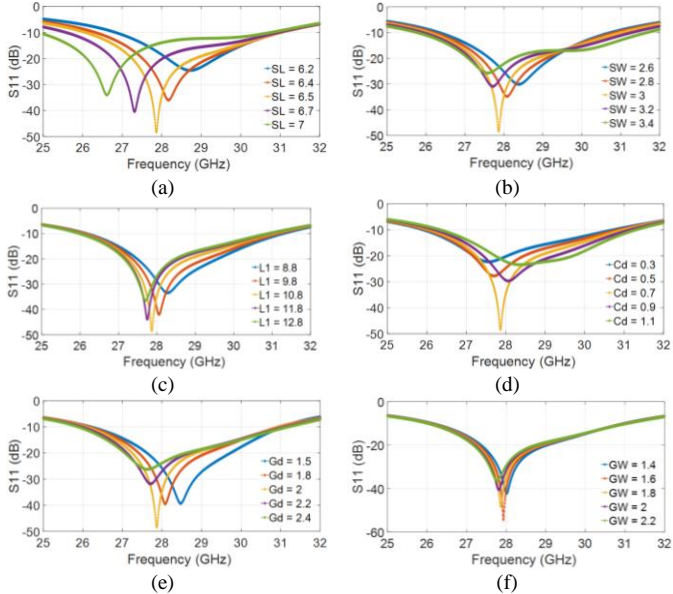

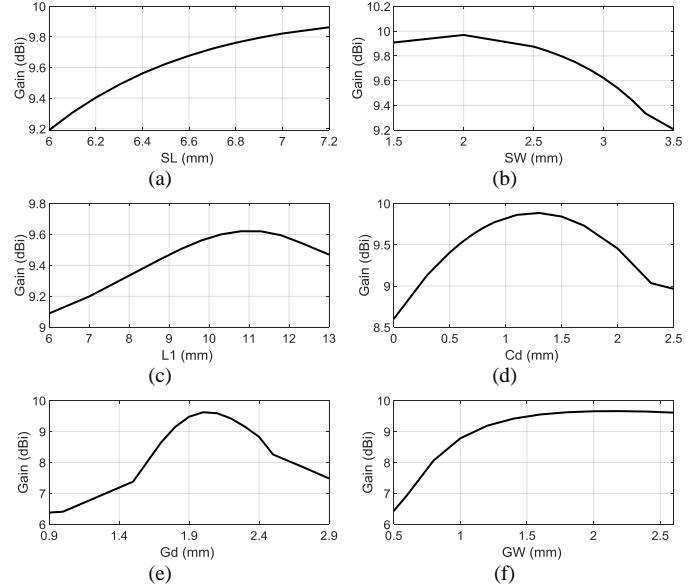

Fig.5. The relationship between the dimensions of the proposed single element antenna and $S_{11}$. (a) $SL$, (b) $SW$, (c) $L1$, (d) $Cd$, (e) $Gd$ and (f) $GW$.[all dimensions are in $mm$]

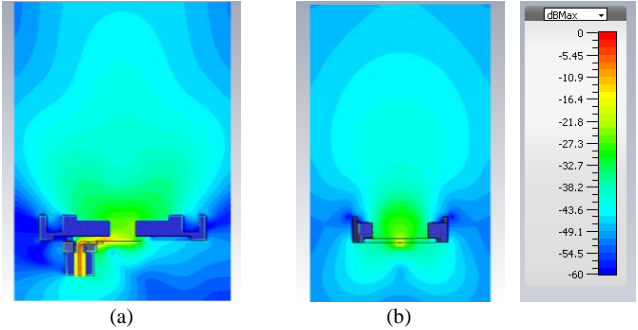

Fig.6. Power flow of the single element antenna at 28 GHz. (a) Cross section of x-z plane and (b) cross section of y-z plane.

Fig.7. The effect of antenna parameters on its directivity at 28 GHz. (a) $SL$, (b) $SW$, (c) $L1$, (d) $Cd$, (e) $Gd$, and (f) $GW$.

## III. OPERATION PRINCIPLES

### A. The Operation Principles of the Single Element Antenna

The proposed antenna consists of a resonant slot surrounded by rectangular cavity and corrugations and it has the same operation principles to proposed antennas in [14] with two main differences as the antennas in [14] are fed using rectangular waveguide structures and fabricated from Aluminum using computer numerical control (CNC) milling machine instead of 3D printing. The resonance frequency of the antenna is mainly controlled by the slot dimensions as the slot resonates at $SL \approx \frac{\lambda}{2}$. Hence, the resonance frequency is inversely proportional to $SL$, while the bandwidth is directly proportional to $SW$ as shown in Fig.5 (a) and (b). The dimensions of the cavity and corrugations have a minor effect on the resonance frequency of the antenna due to the resonance of $TM_{12}$ inside the cavity and $TE$ mode inside the corrugations as explained in details in [14] and shown in Fig.5 (c-f). The antenna has a gain of 9.6 dBi at 28 GHz due to the direct radiation from the slot supported by the radiation from the rectangular cavity and corrugations as shown in Fig.6. For example, the proposed antenna has a gain of ~ 5 dBi at 28 GHz due to the radiation from the slot only and the gain is improved by ~ 3.5 dBi after the addition of the corrugations and by ~ 1.2 dBi after the addition of the recatngualr cavity. Hence, the dimensions of the antenna are numerically optimized to maximize its gain at 28 GHz as shown Fig.7, while keeping a fixed thickness of $t = 3\ mm$ to keep a highly compact and thin profile.

### B. Wall Effect and Real Time Beam Switching Mechanism

An introduction of a 3D printed metallized wall on one side of the antenna as shown in Fig.3, will create asymmetric electric field and asymmetric surface current on the surface of the antenna, which will steer the antenna beam to the opposite direction of the wall in the y-z plane. The beam steer-ability is proportional to the wall height ($Wh$) until reaching the saturation limit. Hence, the maximum steer-ability angle, which could be achieved using this technique is 30° as shown in Fig.8.

The angle at which the beam is steered can be controlled by simply controlling $Wh$. The relationship between the gain of the antenna, beam steer-ability and wall height is described in Table.2. Unlike other beam steering techniques, the proposed sidewall not only able to steer the beam of the antenna, but it is able to increase the directivity and the gain of the antenna. The gain of the antenna is decreased by a maximum of 0.9 dBi once a wall with a height of $4.5\ mm$ is introduced, then the directivity and gain of the antenna start to increase proportionally to the wall height. The directivity and gain of the antenna recovers the losses once the wall height is more than the operation wavelength ($\lambda$); hence the antenna has a gain of 9.6 dBi when there is no wall and a gain of 10.0 dBi when the wall height is $11\ mm\ (\approx \lambda)$. A further increment with the wall height will increase the gain of the antenna until reaching a saturation limit. The saturation limit for the gain is found to be 12.5 dBi, when the wall height is $\approx 5\ \lambda$ as shown in Table.2.

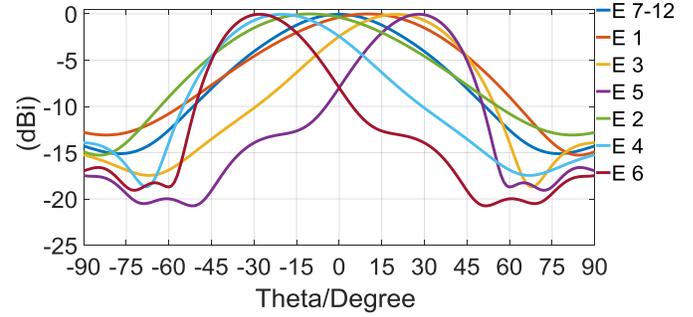

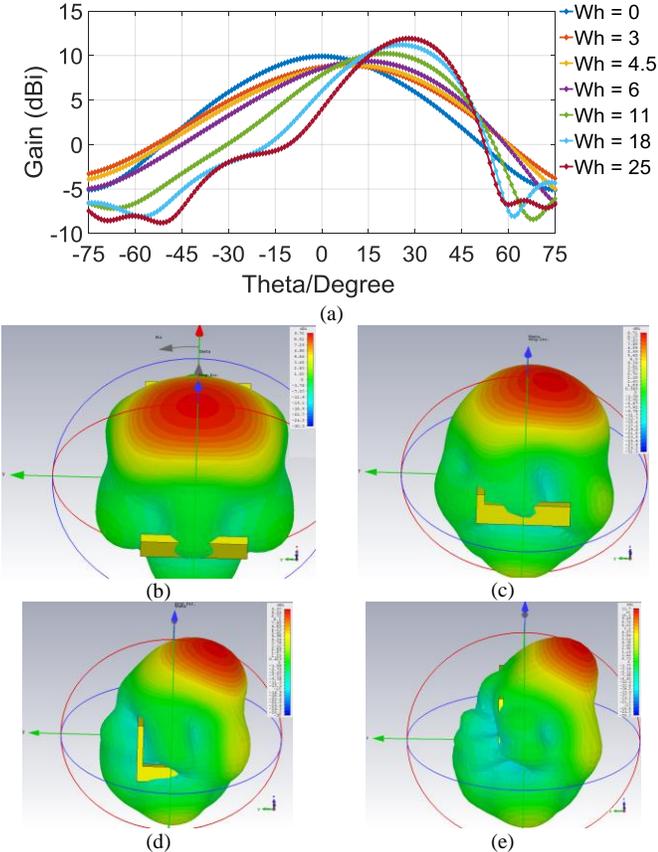

Fig.8. The effect of wall height ($wh$) on the radiation patterns of the antenna. (a) 2D radiation patterns of H-plane (y-z plane) for different wall height in $mm$, (b) 3D radiation patterns of the antenna with no wall $Wh = 0\ mm$, (c) 3D radiation pattern for $Wh = 4.5\ mm$, (d) $Wh = 11\ mm$ and (e) $Wh = 25\ mm$.

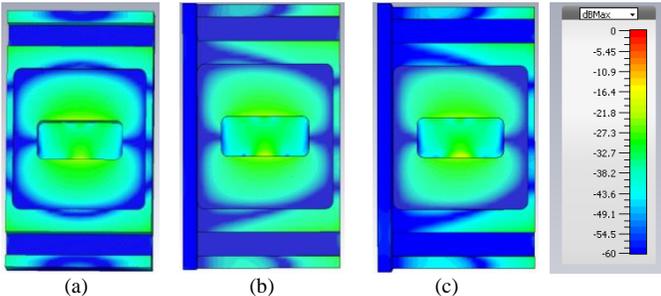

(a)  (b)  (c)

Fig.9. The electric field (E-field) distribution on the antenna surface with/without the presence of the wall. (a) No wall, (b) $Wh = 11\ mm$ and (c) $Wh = 25mm$.

Fig.10. Continuous coverage in the range of $\mp 30°$ in the elevation by using 6 elements with different side wall (Element 1 to 6) and one element that radiate in the boresight direction (any element from 7-12). E in the caption of the graph stands for element.

Table.2. The relationship between wall height ($Wh$), directivity and beam direction at 28 GHz. ($\lambda = 10.71\ mm$)

| Wall Height $Wh\ (mm)$ | Antenna gain $(dBi)$ | Main beam direction (y-z) plane |
|---|---|---|
| No wall 0 | 9.6 | 0° |
| 3 (0.28 $\lambda$) | 8.5 | 7° |
| 4.5 (0.42 $\lambda$) | 8.7 | 10° |
| 6 (0.56 $\lambda$) | 9.1 | 15° |
| 11 ($\lambda$) | 10.0 | 20° |
| 16 (1.5 $\lambda$) | 10.5 | 25° |
| 21.5 (2 $\lambda$) | 11.4 | 27° |
| 25 (2.4 $\lambda$) | 11.7 | 29° |
| 32 (3 $\lambda$) | 12.1 | 30° |
| 43 (4 $\lambda$) | 12.4 | 30° |
| 54 (5 $\lambda$) | 12.5 | 30° |

To gain more understanding on the effect of the wall on the performance of the proposed antenna, the electric field on the antenna surface is studied. As shown in Fig.9, the introduction of the wall has created asymmetric electric field and surface current distribution on the antenna surface, due to reflection of the electromagnetic waves from the wall which is responsible for beam steering. In other words, the wall has pushed the surface wave on the surface of the antenna to the opposite direction of the wall resulting in beam steering and gain enhancement. With presence of no wall, the electromagnetic (EM) energy is coupled to the antenna surface using the slot and it propagates symmetrically on the antenna surface as shown in Fig.9 (a). Hence, part of the EM energy is radiated directly by the slot and part of it is excited by the cavity and the two corrugations; however, the rest of the EM energy is un-radiated and ultimately leaked and lost at antenna edges. Nevertheless, the introduction of the wall on one of the antenna sides will re-direct and reflect the leaked energy at the antenna edge to the surface of the antenna so it can be re-radiated in the opposite direction as shown in Fig.9 (b) and (c), resulting in a directivity and gain enhancement, once the wall is higher than the operation wavelength $\lambda$.

However, using the proposed single element with different sidewall heights in a MIMO configuration will achieve continuous real time coverage in the range of $\mp 30°$ in the elevation plane as shown in Fig.8 and Fig.10. Continuous real time coverage is achieved by switching between different elements that have different sidewall heights and radiate in different direction.

Deploying the single element with fixed and different sidewall heights in a MIMO environment will deliver different beams at different angles which delivers a continuous coverage in the elevation for $\pm \approx 30°$. For example, there are 6 elements radiate in the boresight direction and another 6 elements designed with different wall heights to deliver beam at different directions at $\pm \approx 10°$, $\pm \approx 20°$ and $\pm \approx 30°$ in the proposed $4 \times 3$ MIMO, where real time beam switching can be simply achieved by electronically switching between these elements as shown in Fig.10, to deliver the beam at the required direction. This is simply achievable with no need of phase shifters as the proposed configuration is a MIMO antenna which implies that each of the proposed antenna element in the MIMO will have its own independent RF chain. In addition, the proposed technique is capable of delivering beams at different directions by either increasing the number of MIMO elements with different sidewalls height to deliver beams at the required direction or by varying the height of the wall in the current configuration before manufacturing.

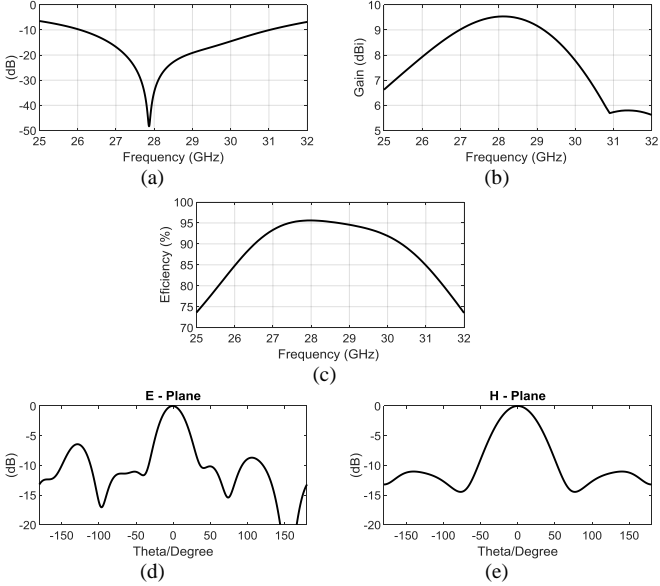

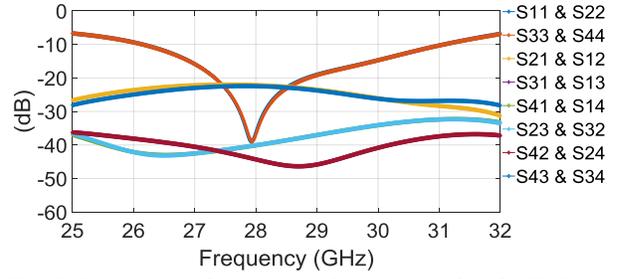

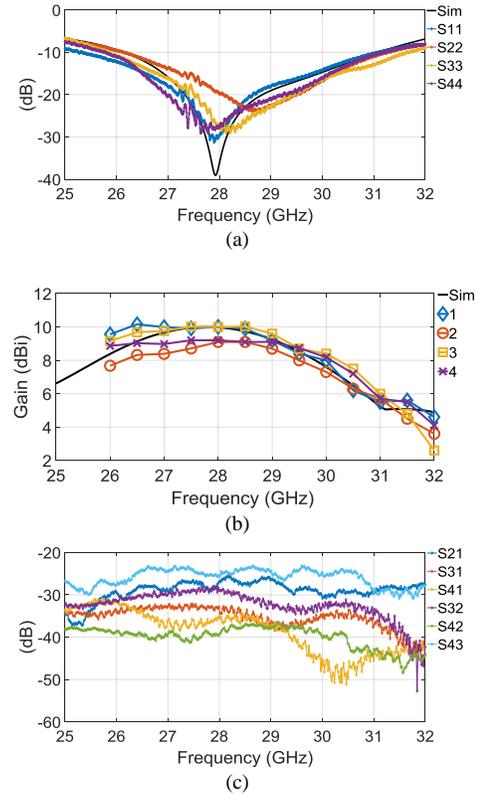

Fig.11. The simulated results of the single element antenna. (a) $S_{11}$, (b) gain, (c) radiation efficiency, (d) E-plane and (e) H-plane.

## IV. RESULTS

### A. Single Element Antenna

The proposed antenna is numerically simulated using CST microwave studio and the simulated reflection coefficient ($S_{11}$), gain, efficiency and radiation patterns are shown in Fig.10. The antenna resonates at 27.9 GHz with a -10 dB simulated bandwidth of 4.9 GHz ranges from 26.1 GHz to 31 GHz as shown in Fig.11 (a). The peak simulated gain of the antenna is 9.6 dBi at 28 GHz and the antenna has a simulated gain of 8 dBi at 26 GHz, rises to 9.17 dBi at 27 GHz and approaching the peak at 28 GHz. Then, the gain drops to 9 dBi at 29.1 GHz and to 8 dBi at 29.9 GHz and to 5.8 dBi at 31 GHz as shown in Fig.11 (b).

The single element antenna has a peak radiation efficiency of 96% at 28 GHz as shown in Fig.11 (c). The radiation efficiency is higher than 85% over the entire bandwidth of the antenna and it is higher than 90% over a bandwidth of 3.8 GHz ranges from 26.5 GHz to 30.3 GHz. The simulated radiation patterns of the antenna are shown in Fig.10 (d), and Fig.11 (e). At 28 GHz, the antenna has a half power beamwdith (HPBW) of 36.7° in the E-plane with side lobe level (SLL) of -9.2 dBi and the HPBW in the H-plane is 56° with SLL of -11 dBi. The front to back ratio (F/B) of the proposed antenna is 6.5 dB. The antenna has a radiation in the boresight direction with good SLL, HPBW performance over a wide bandwidth ranges between 26 GHz and 30.5 GHz. The simulated radiation patterns of the antenna are shown in Fig.10 (d) and Fig.11 (e). At 28 GHz, the antenna has a half power beamwdith (HPBW) of 36.7° in the E-plane with side lobe level (SLL) of -9.2 dBi and the HPBW in the H-plane is 56° with SLL of -11 dBi. The front to back ratio (F/B) of the proposed antenna is 6.5 dB. The antenna has a radiation in the boresight direction with good SLL, HPBW performance over a wide bandwidth ranges between 26 GHz and 30.5 GHz.

### B. 2 × 2 MIMO

Arranging the proposed single element antenna in a 2 × 2 configuration as shown in Fig.3, will introduce a small shift of less than 100 MHz in the resonance frequency of each element in comparison to the single element antenna. Each element in the MIMO resonates at 27.94 GHz with a -10 dB bandwidth of 4.9 GHz ranges between 26.1 GHz and 31 GHz as shown in Fig.12. The simulated mutual coupling (isolation) between any two elements in the MIMO is less than -22 dB for the horizontal adjacent elements (i.e. element 1 and 2) and less than -33dB for the vertical adjacent elements. (i.e. element 1 and 3).

Fig.12. The simulated $S$ parameters of the proposed 2 × 2 MIMO antenna.

Fig.13. The simulated and measured results of the 2 × 2 MIMO. (a) simulated and measured $S_{11}$, (b) simulated and measured gain and (c) measured mutual coupling.

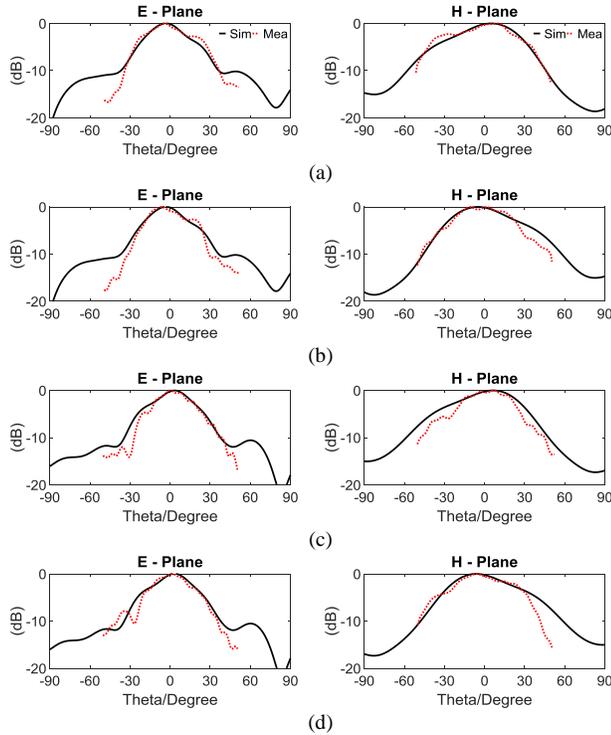

Fig.14. The simulated and measured radiation patterns of the proposed 2 × 2 MIMO at 28 GHz. (a) Element 1, (b) element 2, (c) element 3 and (d) element 4.

Table.3. The summary of far-field characteristics of the 2 × 2 MIMO at 28 GHZ.

| Element | Gain Simulated (dBi) | Gain Measured (dBi) | HPBW E-plane | SLL E-plane | HPBW H-plane | SLL H-plane |
|---|---|---|---|---|---|---|
| 1 | 10.0 | 10 | ~32.4° | -10 | ~55.1° | -14 |
| 2 | 10.0 | 9.1 | ~32.4° | -10 | ~55.1° | -14 |
| 3 | 10.1 | 10.1 | ~34.4° | -11 | ~56.6° | -15 |
| 4 | 10.1 | 9.2 | ~34.4° | -11 | ~56.6° | -15 |

Each element in the 2 × 2 MIMO radiates in the boresight direction with a simulated gain at 28 GHz of 10 dBi for element 1 and 2 and a gain of 10.1 dBi for element 3 and 4. The measured results of the 2 × 2 MIMO are shown in Fig.12, Fig.13 and summarized in Table.3. The gain of the antennas is measured using gain comparison method described in [19] and the radiation patterns are measured using planar near field scanner (NSI) which provide an accurate radiation pattern measurement within a range of $\mp 40° \leq \phi \leq \mp 60°$. A good agreement is found between the simulated and measured bandwidth, gain and radiation patterns with some discrepancies due to combination of fabrication tolerances. In more details, there are three main fabrication tolerances affect the reflection coefficient ($S_{11}$) and the gain performance of each element in the proposed MIMO as following: firstly, the tolerances of the PCB manufacturing process in the feeding layer. Secondly, the fabrication tolerances of the 3D printed antenna as the used 3D printer has a 100 $\mu m$ resolution which has a minor effect on the $S_{11}$ due to the sensitivity of the resonance frequency of the single element to the dimension of the slot as shown in Fig.5 (a) and (b). Thirdly, the tolerances due to misalignment of the PE44968/PE44489 mini-smp connector during the soldering process. In fact, the $S_{11}$ is very sensitive to any vertical or horizontal misalignment in the connector position while soldering it to the transmission line and the mini-SMP pad causing a deviation in the resonance frequency of the antennas. Measured results show that all four elements have a -10 dB bandwidth of ∼ 5GHz with resonance frequency around 28 GHz and an isolation of less than -25 dB between all elements as shown in Fig.13 (a) and (c). In addition, measured results as shown in Fig.14 and Table.3 shows that all elements have good measured gain performance as element 1 and 3 have a measured gain of 10 dBi and 10.1 dBi, while elements 2 and 4 have a measured gain 9.1 dBi and 9.2 dBi at 28 GHz.

*C. 4 × 3 MIMO*

Simulation results of the proposed 4 × 3 MIMO show that all 12 elements resonate at 27.8 ± 0.13 GHz and the simulated mutual coupling between any two elements in the MIMO is less than -26 dB between the elements with the wall and less than -36 dB between any elements with the wall and the closest elements with no walls. Furthermore, the measured reflection coefficients of the six steerable elements with side walls (Element 1 to 6) are shown in Fig.15 (a), while the measured gain performance are shown in Fig.15.(b) and the measured mutual coupling for element 1 and 5 are shown in Fig.15 (c) and (d). Measured results show that all six elements have a measured -10 dB bandwidth of higher than 3.8 GHz covers the frequency band from 26 GHz to 29.8 GHz with resonance frequencies centered around 28 GHz with a shift of less than 500 MHz to higher resonance for some elements. Besides, the measured mutual coupling between any two elements in the MIMO is less than -28 dB as shown in Fig. 15 (c) which presents the measured mutual coupling for element 1 and 5. The measured mutual coupling for all of the rest of the elements in the MIMO are lower than -28 dB and follow similar trend to element 1 and 5. Fig.15 (b) shows the measured gain of the six elements with side wall. All elements with wall have the measured peak gain at 28.5 GHz instead of 28 GHz due to the shift in the reflection coefficient, with gain losses of less than 1 dB in comparison to the simulated results as summarized in Table.4. For example, element 1 and 2 have a peak gain of 11.5 dBi and 10.8 dBi, while element 3 and 4 have a peak gain of 10.7 dBi and 9.6 dBi and the peak gain values of element 5 and 6 are 9.5 dBi and 9.7 dBi. The measured radiation patterns of the six elements with wall are shown in Fig.16 and HPBW values are summarized in Table.4, where generally a good agreement is found between the simulated and measured radiation patterns. Fig.16 shows that the measured beam of element 1 and 2 in the H-plane is steered by +28° and -32° in a similar trend to the simulated performance. However, the measured beam of element 3 and 4 beam are steered by ∼ +25° and ∼ −17° as shown in Fig.16 (c) and (d) with good agreement with the simulated patterns. Finally, the beam of element 5 and 6 have their beams steered by +13° and −11° as shown in Fig.16 (e) and (f).

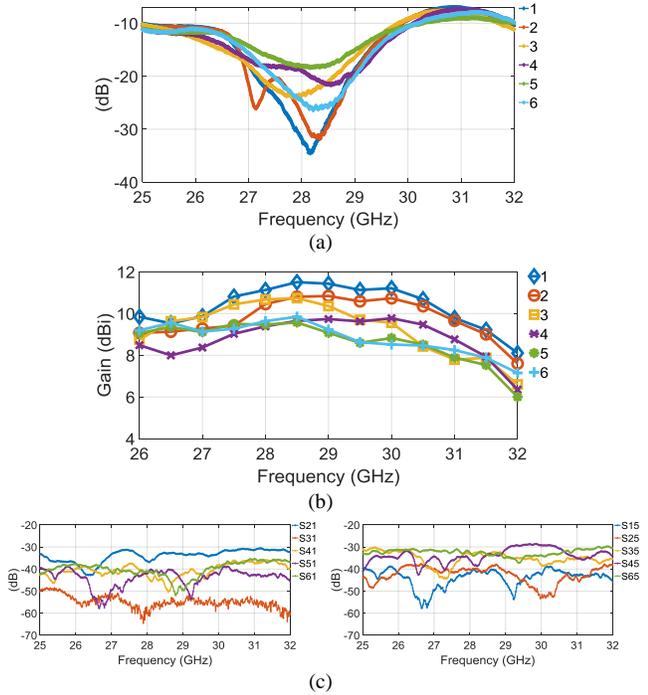

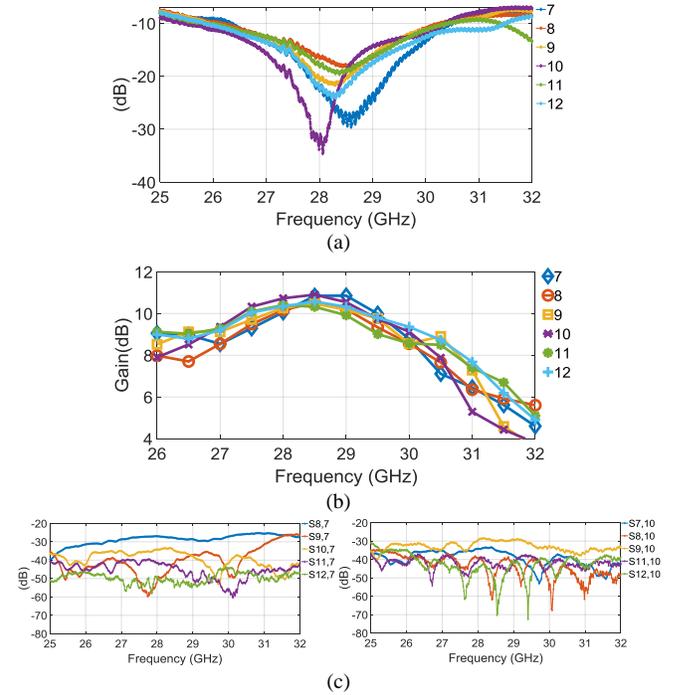

Fig.15. Measured results of the steer-able six elements with wall. (a) Reflection coefficients, (b) gain (c) $S$ parameters for element 1 and 5.

Fig.17. The measured results of the six elements that radiate in the boresight direction. (a) Reflection coefficients and (b) gain (c) $S$ parameters for element 7 and element 10.

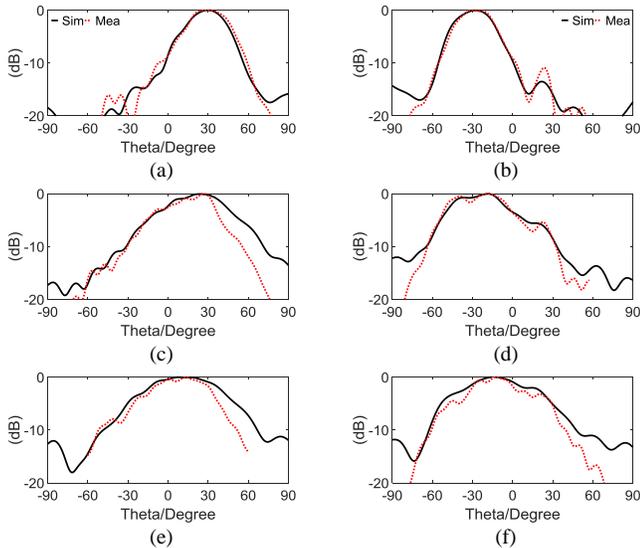

Fig.16. The simulated and measured radiation patterns of the steerable elements at 28 GHz in the (y-z) plane. (a) Element 1, (b) element 2, (c) element 3, (d) element 4, (e) element 5 and (f) element 6.

The measured reflection coefficients, mutual coupling and gain of the six elements that radiate in the boresight direction (Element 7 to 12) are shown in Fig.17 and Fig.18. All six elements have a bandwidth of $\approx$ 4 GHz and they resonate at 28 GHz with a shift of less than 350 MHz to higher resonance frequency for element 7, 8 and 9 and a shift of $\approx$ 500 MHz for element 8 and $\approx$ 600MHz for element 7. All elements have a measured gain of higher than 10 dBi at 28 GHz and the peak gain of $\approx$10.5 dBi is observed at 28.5 GHz for most elements as shown in Fig.17 (b). The measured radiation patterns of Element 7 to 12 are shown in Fig.18 and they show that all six elements radiate in the boresight direction with a measured HPBW of $\approx 25° \mp 5°$ in the E-plane and HPBW of $\approx 45° \mp 5°$ in the H-plane at 28 GHz as summarized in Table.5, with low SLL performance and a good agreement with the simulated patterns.

Table.4: Summary of gain and HPBW of the steer-able elements with wall in the proposed 4 × 3 MIMO at 28 GHz.

| Element | Gain Simulated 28 GHz (dBi) | Gain Measured 28 GHz (dBi) | Gain Peak 28.5 GHz (dBi) | HPBW H-plane Measured | HPBW E-plane Measured |
|---|---|---|---|---|---|
| 1 | 11.75 | 11.1 | 11.5 | 33° | 36.5° |
| 2 | 11.7 | 10.7 | 10.8 | 34° | 38° |
| 3 | 10.4 | 10.6 | 10.7 | 49.6° | 35.9° |
| 4 | 10.3 | 9.3 | 9.6 | 52.3° | 41° |
| 5 | 9 | 9.4 | 9.5 | 54.5° | 42.5° |
| 6 | 9.5 | 9.5 | 9.7 | 44.4° | 45.6° |

Table.5. Summary of gain and HPBW of the boresight elements in the proposed 4 × 3 MIMO at 28 GHz.

| Element | Gain Simulated 28 GHz (dBi) | Gain Measured 28 GHz (dBi) | Gain Peak 28.5 GHz (dBi) | HPBW H-plane Measured | HPBW E-plane Measured |
|---|---|---|---|---|---|
| 7 | 10.7 | 10.1 | 10.8 | 41° | 29° |
| 8 | 10.4 | 10.2 | 10.5 | 47.8° | 28.4° |
| 9 | 10.2 | 10.3 | 10.5 | 51.9° | 27.8° |
| 10 | 11 | 10.7 | 10.8 | 39.5° | 21° |
| 11 | 10.1 | 10.4 | 10.3 | 46.7° | 28.3° |
| 12 | 10.5 | 10.2 | 10.5 | 48.3° | 25.5° |

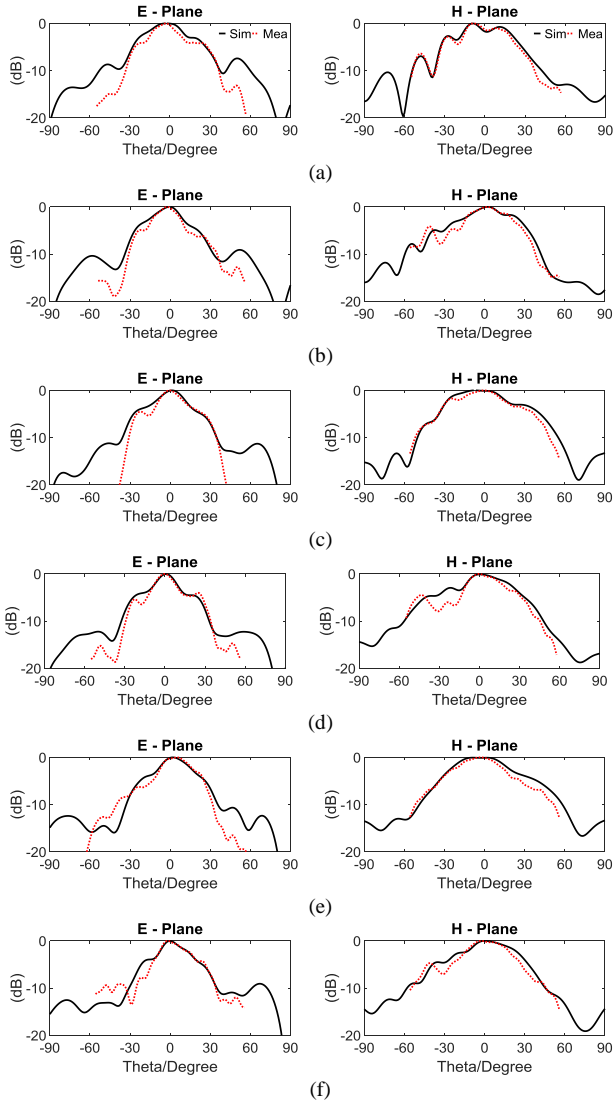

Fig.18. The simulated and measured radiation patterns of the elements that radiate in the boresight direction. (a) Element 7, (b) element 8, (c) element 9, (d) element 10, (e) element 11 and (f) element 12.

## V. Diversity Analysis

Several diversity parameters have been implemented to caclculate the diversity performance of the proposed MIMO antennas based on the MIMO far-field characteristics. Those include: envelop correlation coefficient ($\rho_e$), effective diversity gain (EDG) and mean effective gain (MEG). The definition and calculation procedure of the respective parameters will not be discussed here, and they can be found in [20-22] for $\rho_e$ and in [23,24] for EDG and in [25,26] for MEG.

Fig.18 shows $\rho_e$ and EDG performance for the propsoed $2 \times 2$ MIMO. Fig.18 (a) shows that all elements in the $2 \times 2$ MIMO have very low and favorable $\rho_e$ performance of less than -50 dB over the entire band. Furthermore, Fig.19 (b) shows that the EDG values between any two elements in the proposed MIMO is higher than 9.99 dB which is close to the maximum 10 dB as discussed in [23].

Fig.20 shows the $\rho_e$ and EDG of element 2 in the proposed $4 \times 3$ MIMO. Fig.19 (a) shows that $\rho_e$ between element 2 and all other elements in the MIMO are less than -50 dB and EDG is higher than 9.97 dB for all elements. All other 11 elements in the $4 \times 3$ MIMO have a similar $\rho_e$ and EDG performance to element 2 with $\rho_e$ being less than -50 dB and with EDG beign higher than 9.97 dB. Finally, the MEG values for all elements in both of the proposed $2 \times 2$ and $4 \times 3$ MIMO are ≈ -3.1 dB at the operation frequency.

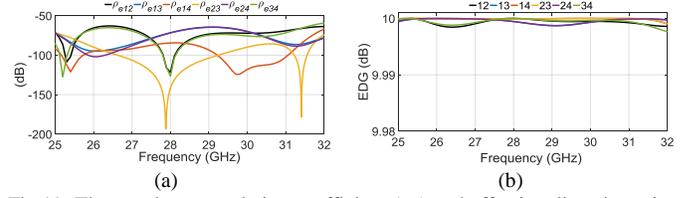

Fig.19. The envelope correlation coefficient ($\rho_e$) and effective diversity gain (EDG) of $2 \times 2$ MIMO. (a) $\rho_e$ and (b) EDG.

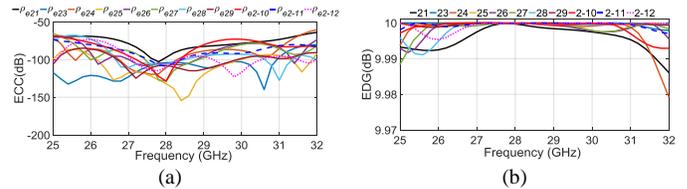

Fig.20. ($\rho_e$) and EDG of element 2 in the $4 \times 3$ MIMO. (a) $\rho_e$ and (b) EDG.

## VI. Conclusion

A design of two novel and low cost MIMO antennas for 5G mm-wave communication are discussed in this paper. The proposed MIMO are fabricated using 3D printing and they operate on the 28 GHz 5G band with a wide bandwidth performance. The first MIMO antenna consists of four elements arranged in $2 \times 2$ configuration that is able to provide radiation in the boresight direction and the second MIMO antenna is able to provide beam steering capabilities as it consists of twelve radiating elements organized in $4 \times 3$ configuration.

The $4 \times 3$ steerable MIMO consists of six elements that provide radiation the boresight direction and another six elements that provide steer-able beam in the elevation plane. The beam of the $4 \times 3$ MIMO is steered mechanically through introducing a metallic wall with different height on the side of the radiating single element structure. The sidewall creates asymmetric electric field on the surface of the antenna, which reflects the beam of the antenna to the opposite direction. The proposed sidewall is able to steer of the beam of the MIMO up to 30° in the elevation plane. Finally, the performance of the proposed MIMO antennas are measured and found to operate as predicted by the numerical simulation tool.